\documentclass[lettersize,journal]{IEEEtran}
\usepackage{amsmath,amsfonts}
\usepackage{algorithmic}
\usepackage{algorithm}
\usepackage{array}
\usepackage[caption=false,font=normalsize,labelfont=sf,textfont=sf]{subfig}
\usepackage{textcomp}
\usepackage{stfloats}
\usepackage{url}
\usepackage{verbatim}
\usepackage{graphicx}
\usepackage{amssymb}
\usepackage{cite}

\begin{document}

\title{Near-Field Multiuser Beam Training for XL-MIMO: An End-to-End Interference-Aware Approach with Pilot Limitations}

\author{Xinyang Li, Songjie Yang, Xiang Ling,~\IEEEmembership{Member,~IEEE,} Jianhui Song, Yibo Wang, Hua Chen,~\IEEEmembership{Senior Member,~IEEE}
\thanks{Xinyang Li, Songjie Yang, Xiang Ling, and Jianhui Song are with the National Key Laboratory of Wireless Communications, University of Electronic Science and Technology of China, Chengdu 611731, China (e-mail: lixy829@std.uestc.edu.cn; yangsongjie@std.uestc.edu.cn; xiangling@uestc.edu.cn; stihao321@gmail.com).}%
\thanks{Yibo Wang is with the Department of Engineering, University of Cambridge, Cambridge CB2 1PZ, U.K. (e-mail: yw755@cam.ac.uk).}%
\thanks{Hua Chen is with the Faculty of Electrical Engineering and Computer Science, Ningbo University, Ningbo 315211, China (e-mail: kchenhua0714@hotmail.com).}%
\thanks{This work has been submitted to the IEEE for possible publication. Copyright may be transferred without notice, after which this version may no longer be accessible.}%
}


\maketitle

\begin{abstract}
Near-field propagation in extremely large-scale MIMO (XL-MIMO) enlarges the beam training (BT) search space by introducing an additional range dimension, which makes conventional codebook-based beam sweeping prohibitively expensive under limited pilot resources, especially for multiuser sub-connected hybrid architectures. This letter proposes a deep-learning-based interference-aware multiuser BT framework (DL-IABT) that directly predicts analog beam indices from a small number of uplink sensing measurements. By exploiting a subarray-level approximation, a far-field codebook is adopted to represent each subarray response with negligible mismatch. To enable end-to-end (E2E) learning, we derive a variant-MSE surrogate loss by eliminating the digital precoder through a closed-form MMSE solution from KKT conditions, which implicitly accounts for multiuser interference (MUI). The proposed network integrates a complex-valued sensing front-end, a shared complex-valued encoder, a Transformer-based multiuser predictor, and a scalable Gumbel--Softmax beam selection head. Simulation results show that DL-IABT achieves near-optimal sum-rate performance while providing markedly higher effective throughput under pilot overhead constraints.
\end{abstract}

\begin{IEEEkeywords}
Extremely large-scale MIMO (XL-MIMO), near-field communications, sub-connected hybrid beamforming, multiuser beam training, deep learning, Transformer, Gumbel--Softmax.
\end{IEEEkeywords}

\section{Introduction}
\IEEEPARstart{T}{he} rapid growth of mobile data traffic has driven wireless systems toward higher carrier frequencies to exploit wider bandwidths. Meanwhile, the shorter wavelength enables the integration of densely packed antenna arrays within compact form factors. XL-MIMO systems equipped with extremely large aperture arrays (ELAAs) exploit near-field spherical wavefronts to achieve fine-grained energy focusing and enhanced beamforming gain \cite{cui2022channel,10934792, wang2024tutorial}. 
However, in the near-field region, the channel must be characterized over both angular and range dimensions, causing the BT search space to grow rapidly with both resolution. 
Moreover, in multiuser XL-MIMO systems, MUI and limited pilot resources further complicate beam selection. 
To reduce hardware complexity and alleviate the training dimensionality, sub-connected hybrid architectures partition the array into subarrays and provide an important structural solution for ELAA systems \cite{yang2024near}.

\IEEEpubidadjcol
In hybrid beamforming (HB) systems, conventional BT schemes typically decouple the design of analog and digital beamformers \cite{11346973}. In the first stage, codebook beams are scanned through hierarchical \cite{xiao2018enhanced, shi2024spatial}, exhaustive, or sampling-based search, and beam selection is performed based on user feedback. In the second stage, the equivalent baseband channel is estimated, and the digital beamformer is obtained. 
For near-field scenarios, recent studies further explore hierarchical codebooks defined in the polar domain to capture both angular and range characteristics \cite{shi2024spatial}. 
However, fundamentally, codebook-based search follows a fixed channel sensing strategy with limited flexibility and low pilot efficiency. 
Moreover, in sub-connected architectures with multi-beam training, the number of feasible beam combinations grows exponentially with the number of subarrays. 
The resulting prohibitive complexity and latency of joint cross-subarray search have motivated various linear-search-based solutions \cite{li2023hierarchical}. 
Nevertheless, such stage-wise designs are not necessarily optimal in multiuser scenarios, since the analog beams selected in the first stage mainly maximize single-user beam gain without directly optimizing system-level performance. When inter-user channel correlation increases or the scattering environment becomes more complex, the selected analog beams may restrict the interference mitigation capability of subsequent digital precoding and lead to reduced sum-rate performance.

In recent years, deep learning has been widely applied to HB and BT in both far-field and near-field systems due to its strong representation capability. 
However, most existing works assume fully digital architectures \cite{park2024end}, fully connected hybrid structures \cite{wang2024learning}, or purely analog designs \cite{nie2025near}, which cannot accommodate the block-diagonal constraints and limited pilot resources inherent in sub-connected architectures. 
Moreover, many E2E learning approaches for beamforming require the network to jointly output analog and digital beamforming matrices \cite{gao2022data}, which significantly increases model size and training difficulty, while methods that focus only on analog beam selection \cite{nie2025near} often fail to adequately suppress E2E under limited pilot conditions. 

This letter proposes an E2E deep neural network for XL-MIMO systems with sub-connected HB architectures, which departs from conventional stage-wise BT and seeks beam selections that better suppress inter-user interference and maximize system sum rate. We adopt a surrogate variant-MSE loss and exploit KKT conditions to decouple digital and analog beamformers. By minimizing the variant-MSE through deep learning, the network learns analog beam indices that implicitly account for E2E mitigation. To improve pilot efficiency, a complex-valued convolutional sensing module is employed to extract latent channel features, which are then processed by a complex-valued precoding layer and a Transformer-based predictor for interference-aware beam prediction and rate enhancement, thereby reducing signal reconstruction error. To address the exponential growth of beam combination candidates with the number of users, we further introduce parameter-shared multi-head outputs with Gumbel–Softmax relaxation, which avoids the backpropagation issues of one-hot encoding and reduces the output dimension of beam prediction to scale linearly with the number of users.

\section{System Model}
\begin{figure}[!t]
\centering
\includegraphics[width=2.5in]{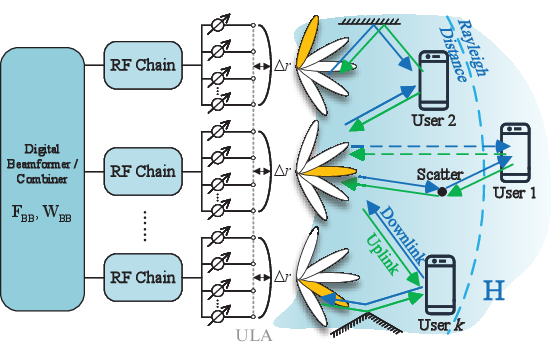}
\caption{Schematic of the XL-MISO sub-connected HB system serving users in mixed near- and far-field regions.}
\label{fig:system}
\end{figure}

We consider a narrowband multiuser XL-MISO system with a sub-connected HB architecture operating in time-division duplex (TDD) mode, serving $K$ single-antenna users, as shown in Fig.~\ref{fig:system}. The base station is equipped with $N_{\rm sub}$ subarrays, each consisting of a half-wavelength-spaced ULA with $N_a$ consecutive antenna elements, while adjacent subarrays are also separated by half-wavelength spacing.

During the uplink sensing stage, at pilot slot $m$, the BS applies a combining matrix $\mathbf{\Phi}^{(m)} \in \mathbb{C}^{N_{\rm sub} N_a \times K}$ to receive uplink pilots with unit average power from the users. After $M$ sensing slots, the received pilot matrix can be written as
\begin{equation}
  \mathbf{Y}_{\rm UL} = \mathbf{\Phi}^H \mathbf{H} + \tilde{\mathbf{N}}_{\rm UL} \in \mathbb{C}^{MK \times K}, 
\end{equation}
where $\mathbf{\Phi} = \left[\mathbf{\Phi}^{(1)}, \dots, \mathbf{\Phi}^{(M)} \right]$; $\mathbf{H} = \left[\mathbf{h}_1, \dots, \mathbf{h}_K \right] \in \mathbb{C}^{N_{\rm sub}N_a \times K}$ denotes the uplink channel matrix, and $\mathbf{h}_k$ is the channel of user $k$; $\tilde{\mathbf{N}}_{\rm UL} = [{\bf\Phi}^{(1)} {\bf N}^{(1)}, \dots, {\bf\Phi}^{(M)} {\bf N}^{(M)}]^T$ is the aggregated uplink noise, where ${\bf N}^{(m)} \in \mathbb{C}^{N_{\rm sub} N_a \times K}$ is the uplink noise matrix of slot $m$ whose entries are i.i.d. complex Gaussian random variables distributed as $\mathcal{CN}(0, \sigma_n^2)$. 

Users are assumed to lie in both near- and far-field regions of the array, and their channels follow spherical-wave propagation. The uplink channel of user $k$ is modeled by the geometric Saleh–Valenzuela model with $L$ dominant paths, 
\begin{equation}
  \mathbf{h}_k = \sqrt{\frac{N_{\rm sub}N_a}{L}} \sum_{\ell=1}^{L} \alpha_{\ell, k} e^{-j\frac{2\pi }{\lambda} r_{\ell, k}} \mathbf{b}(\theta_{\ell, k}, r_{\ell, k}), 
\end{equation}
where $\alpha_{\ell, k}, \theta_{\ell, k}$, and $r_{\ell, k}$ denote the complex gain, angle of arrival, and distance from the array center to the user (or scatterer), respectively.
The near-field array response vector is
\begin{equation}
  \begin{aligned}
  \mathbf{b}(\theta, r) = \frac{ \Big[ e^{-j\frac{2\pi}{\lambda} \left(r^{(1)} - r\right)}, \dots, e^{-j\frac{2\pi}{\lambda} \left(r^{(N_{\rm sub}N_a)} - r\right)} \Big]^T}{\sqrt{N_{\rm sub}N_a}} 
  \end{aligned}
\end{equation}
where $r^{(n)} = \sqrt{r^2 + \delta_n^2 d^2 - 2 r \delta_n d \sin\theta}$ is the distance between antenna element $n$ and the user (or scatterer), with $\delta_n = (2n - N_{\rm sub} N_a - 1) \big/ 2, n=1, \dots, N_{\rm sub} N_a.$

By exploiting channel reciprocity in TDD systems \cite{smith2004direct}, the downlink channel is given by $\mathbf{H}^H$. In the downlink transmission stage, the BS employs a block-diagonal analog beamformer $\mathbf{F}_{\rm RF}={\rm diag} \left\{\mathbf{f}_1, \dots, \mathbf{f}_{N_{\rm sub}} \right\} \in \mathbb{C}^{N_{\rm sub} N_a\times N_{\rm sub}}$, where $\mathbf{f}_n \in \mathbb{C}^{N_a \times 1}$ is the steering vector applied by subarray $n$, and a digital beamformer $\mathbf{F}_{\rm BB}= \left[ \mathbf{f}_{{\rm BB}, 1}, \dots, \mathbf{f}_{{\rm BB}, K} \right] \in \mathbb{C}^{N_{\rm sub} \times K}$, The received downlink signal is
\begin{equation}
  \mathbf{y}_{\rm DL} = \mathbf{H}^H \mathbf{F}_{\rm RF} \mathbf{F}_{\rm BB} \mathbf{s} + \mathbf{n}_{\rm DL} \in \mathbb{C}^{K \times 1}, 
\end{equation}
where $\mathbf{s} = [s_1, \dots, s_K]^T$ satisfies $\mathbb{E}[\mathbf{ss}^H]=\mathbf{I}_K$, and $\mathbf{n}_{\rm DL}\sim\mathcal{CN}(\mathbf{0}, \sigma_n^2 \mathbf{I}_K)$.

For multiuser multi-beam training, each subarray steering vector is selected from a codebook $\mathcal{F}$. The joint beam selection and digital precoder optimization problem for sum-rate $R_{\rm sum}$ maximization is formulated as
\begin{equation}
  \begin{aligned}
    \underset{\mathbf{F}_{\rm RF}, {\bf F}_{\rm BB}}{\rm maximize} \quad &\sum_{k=1}^K \log_2 \left( 1 + \frac{ \left| \mathbf{h}_k^H \mathbf{F}_{\rm RF} \mathbf{f}_{{\rm BB}, k} \right|^2}{\sum_{i \neq k} \left| \mathbf{h}_k^H \mathbf{F}_{\rm RF} \mathbf{f}_{{\rm BB}, i} \right|^2 + \sigma_n^2} \right), \\
    {\rm subject\ to} \quad &{\bf f}_n\in\mathcal{F},\quad n=1, \dots, N_{\rm sub},\\
    &\left\lVert \mathbf{F}_{\rm RF} \mathbf{F}_{\rm BB} \right\rVert_F^2 \le P_t,
  \end{aligned}\label{SumRateMax}
\end{equation}
where $P_t$ denotes the maximum transmit power.

\section{Pilot-Efficient E2E Multiuser Beam Training}

The optimal HB design in \eqref{SumRateMax} requires a joint search over all possible analog beam combinations in the codebook $\mathcal F$, whose training overhead grows exponentially with the number of subarrays $N_{\rm sub}$. The latest work \cite{zhang2019subarray} on multiuser BT under sub-connected architectures mainly focuses on single-user rate maximization and seldom consider the impact of E2E on system sum rate. 
To enable efficient beam prediction, we first exploit the structural property of sub-connected arrays.

\cite{yang2024near} proposes to partition a large ULA into a sufficient number of subarrays and assign different far-field steering vectors to different subarrays. In this way, the overall array steering vector can approximate the near-field steering vector, i.e.,
\begin{equation}
  \mathbf{g}(\theta, r) \big|_{d_{\rm cen}=N_ad} = \mathbf{b}(\theta,r) \approx \left[ \mathbf{a}_{N_a}^T(\theta_1), \dots, \mathbf{a}_{N_a}^T(\theta_{N_{\rm sub}}) \right]^T,
\end{equation}
where $d_{\rm cen}$ denotes the spacing between subarray centers, $\theta_n$ is the local incident angle determined by the spherical-wave geometry, and the far-field array steering vector $\mathbf{a}_n(\theta) = {1}/{\sqrt n} [1, e^{j\pi\sin\theta}, \dots, e^{j\pi(n-1)\sin\theta} ]^T$. 
For a fixed propagation distance $r$, this approximation becomes more accurate as the number of subarrays increases. Each subarray has an aperture $D_{\rm sub} = (N_a - 1) d$. The upper bound of the maximum path difference between antenna elements and a user (or scatterer) is $\Delta r \approx D^2_{\rm sub} \big/ (8r)$, which yields a maximum phase error $\Delta \phi \approx 2\pi \Delta r / \lambda$. When $\Delta \phi < \pi/8$, the plane-wave approximation error can be regarded as negligible \cite{sherman1962properties}. 

Assume that the ULA covers $(-\pi/2, \pi/2]$. The beam angles in the far-field codebook are obtained by uniform quantization in the spatial-frequency domain, i.e., $\sin\hat{\theta}_i=-1+2i/N_q,\ i=1,\dots,N_q$, where $N_q$ denotes the codebook size.
Accordingly, the far-field beam codebook applicable to each subarray is given by $\mathcal{F} = \{\mathbf{a}_{N_a}(\hat{\theta}_i)\mid i=1, \dots, N_q \}$.



\subsection{MMSE-Based Training Objective}

Directly maximizing the sum-rate in \eqref{SumRateMax} is difficult in a learning framework due to its nonconvex form and the discrete codebook constraint. To obtain a tractable training objective, we eliminate the digital beamformer via a closed-form MMSE solution \cite{shi2011iteratively}.

For a given analog beamformer $\mathbf F_{\rm RF}$ and channel $\mathbf H$, the optimal linear digital precoder that minimizes the signal reconstruction error can be written as
\begin{equation}
  \mathbf F_{\rm BB}^\star = \arg\min_{\mathbf F} \mathbb{E} \left[ \left\lVert \mathbf{s} - \beta^{-1} \mathbf{y}_{\rm DL} \right\rVert_2^2 \right], \label{FbbOptProb}
\end{equation}
where the scaling factor $\beta\in\mathbb{R}_+$ is introduced to align the amplitude of the effective received signal with the source $\mathbf s$, thereby facilitating optimization under the total power constraint. The solution of \eqref{FbbOptProb} is given by the closed form derived from KKT conditions as
\begin{equation}
  \mathbf F_{\rm BB}^\star = \beta \left( \mathbf F_{\rm RF}^H \mathbf H\mathbf H^H \mathbf F_{\rm RF}+\frac{K\sigma_n^2}{P_t} \mathbf{F}_{\rm RF}^H \mathbf{F}_{\rm RF}\right)^{-1} \mathbf{F}_{\rm RF}^H \mathbf{H}. \label{FbbClosedForm}
\end{equation}
Substituting $\mathbf F_{\rm BB}^\star$ into the active power constraint yields the closed-form scaling factor:
\begin{equation}
\beta = \sqrt{{P_t}\Big/{{\rm tr} \left( {\bf F}_{\rm RF}\tilde{\bf F}_{\rm BB} \tilde{\bf F}_{\rm BB}^H  {\bf F}_{\rm RF}^H\right)}}. \label{Beta}
\end{equation}
Then, substituting \eqref{FbbClosedForm} and \eqref{Beta} into \eqref{FbbOptProb} leads to the following variant-MSE objective
\begin{equation}
  \begin{aligned}
    \mathcal L(\mathbf F_{\rm RF}) = {\rm tr}\bigg( \Big({\bf I}_K + \frac{P_t}{K\sigma_n^2}{\bf H}^H{\bf F}_{\rm RF}&({\bf F}_{\rm RF}^H{\bf F}_{\rm RF})^{-1} \\
          &\quad\times {\bf F}_{\rm RF}^H{\bf H} \Big)^{-1} \bigg),
  \end{aligned}\label{Loss}
\end{equation}
which implicitly captures E2E suppression. This loss decreases when the effective channel matrix becomes better conditioned, which is known to correlate with higher multiuser sum rate. Based on this formulation, the original combinatorial optimization in \eqref{SumRateMax} is transformed into learning a mapping, which predicts analog beam indices from limited uplink sensing measurements $[i_1, \dots, i_{N_{\rm sub}}] = \mathcal G_\theta(\mathbf Y_{\rm UL})$, where $i_n \in \{1, \dots, N_q \}$. The analog beamformer can be expressed by
\begin{equation}
  \hat{\mathbf{F}}_{\rm RF} = {\rm diag} \left\{ \mathbf{a}_{N_a} \left(\hat{\theta}_{i_1}\right), \dots, \mathbf{a}_{N_a}\left(\hat{\theta}_{i_{N_{\rm sub}}}\right) \right\}.
\end{equation}
The digital precoder is then obtained by the closed-form MMSE solution above. This decoupling enables E2E training while explicitly accounting for E2E, and avoids the need to jointly learn both analog and digital beamformers.

\subsection{Deep-Learning-Based Interference-Aware Multiuser Beam Training (DL-IABT)}
\begin{figure}[!t]
\centering
\includegraphics[width=3in]{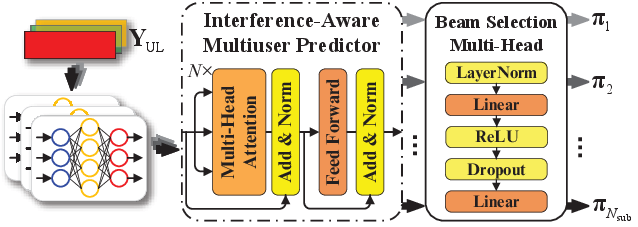}
\caption{Architecture of the proposed interference-aware multiuser beam prediction network.}
\label{fig:network}
\end{figure}

Based on the MMSE-based objective, we design a task-oriented network that predicts the analog beam indices directly from uplink sensing measurements $\mathbf Y_{\rm UL}$. As illustrated in Fig.~\ref{fig:network}, the proposed architecture consists of a complex-valued sensing front-end, a shared feature encoder, an interference-aware multiuser predictor, and a scalable beam selection head.

\subsubsection{Complex-valued sensing front-end}

The proposed network is trained to predict analog beam indices from uplink sensing observations. To integrate the physical sensing interface into E2E learning, we parameterize the uplink measurement process by a bias-free complex-valued grouped convolution layer. During training, each channel realization $\mathbf{H}^{(s)}$ is first broadcast along the measurement dimension to form a tensor $\tilde{\bf H}^{(s)} \in \mathbb{C}^{N_{\rm sub}N_a \times M \times K}$ with $\tilde{\bf H}_n^{(s)} = {\bf H}^{(s)}$. We then apply noise injection as data augmentation by superimposing an i.i.d. AWGN tensor $\tilde{\bf N}^{(s)}$ with its elements following $\mathcal{CN}(0, \sigma_n^2)$, yielding $\tilde{\bf H}^{(s)}_{\rm noisy} = \tilde{\bf H}_n^{(s)} + \tilde{\bf N}^{(s)}$, which is used only in training to improve robustness. 

The grouped convolution implements per-slot linear projections that emulate time-domain beam switching. Specifically, the kernel group associated with slot $n$ learns a sensing matrix ${\bf \Phi}^{(n)}$, and the resulting virtual measurement satisfies
\begin{equation}
  \tilde{\bf Y}_{\rm UL} = f_{\bf\Phi} \left( {\bf H}, \tilde{\bf N}^{(s)} \right) \triangleq \mathcal{G}_{\bf\Phi}(\mathbf{H}) + \tilde{\bf N}_{\rm UL}, 
\end{equation}
where $\mathcal{G}_{\bf\Phi}(\cdot)$ denotes the linear projection induced by grouped convolutions. To preserve the linear sensing model, all bias terms are disabled, and the complex convolution is realized by two parallel real-valued grouped convolutions for the real and imaginary kernels, respectively.

In the training phase, $\mathbf{H}$ is used to evaluate the loss but is not fed into downstream modules beyond the sensing front-end; in the inference phase, the network operates only on the measurement matrix $\mathbf{Y}_{\rm UL}$ which is obtained by fixing the learned combining beamformer $\bf\Phi^\star$.

\subsubsection{Shared complex-valued feature encoder}
For neural processing, we reshape $\mathbf Y_{\rm UL}$ into a tensor $\bar{\mathbf Y}_{\rm UL}\in\mathbb C^{K\times M\times K}$, where the first dimension indexes users, the second indexes the $M$ sensing slots, and the third corresponds to the $K$ pilot streams.

For each user $k$, we take the slice $\mathbf Z_k=\bar{\mathbf Y}_{\rm UL}(k,:,:)\in\mathbb C^{M\times K}$, and map it to a compact embedding through a shared complex-valued MLP:
\begin{equation}
\mathbf e_k^{(\ell+1)}=
\sigma\!\left({\rm CBN}\!\left(\mathbf W^{(\ell)}\mathbf e_k^{(\ell)}+\mathbf b^{(\ell)}\right)\right),
\quad \ell=1,\dots,L_{\rm enc},
\end{equation}
where $\mathbf e_k^{(1)}={\rm vec}(\mathbf Z_k)$,
$\mathbf W^{(\ell)}\in\mathbb C^{d_{\ell+1}\times d_\ell}$ and $\mathbf b^{(\ell)}\in\mathbb C^{d_{\ell+1}}$ are learnable parameters,
${\rm CBN}(\cdot)$ denotes complex batch normalization,
and $\sigma(\cdot)$ is a complex-valued ReLU function.
The final embedding is $\mathbf e_k=\mathbf e_k^{(L_{\rm enc}+1)}\in\mathbb C^{d}$.

All users share the same encoder parameters $\{\mathbf W^{(\ell)},\mathbf b^{(\ell)}\}$, yielding $\mathbf E=[\mathbf e_1,\dots,\mathbf e_K]^T\in\mathbb C^{K\times d}$. This shared encoder preserves phase information in the sensing measurements and provides compact user features for subsequent interference-aware prediction.

\subsubsection{Interference-aware multiuser predictor}

To capture user coupling, we employ a multi-layer self-attention encoder operating across the user dimension.
The complex embeddings $\mathbf E\in\mathbb C^{K\times d}$ are first converted into real-valued tokens $\mathbf X_0
 = \big[\Re(\mathbf E),\ \Im(\mathbf E)\big] \in\mathbb R^{K\times d_{\rm m}}$, where $d_{\rm m}=2d$. The tokens are processed by $L_{\rm tfm}$ stacked Transformer encoder layers.
Let $\mathbf X_\ell$ denote the input of layer $\ell$.
Each layer performs multihead self-attention followed by a feedforward mapping,
\begin{align}
\mathbf{Z}_\ell = \mathbf{X}_\ell + {\rm MHA}_\ell \left( {\rm LN}(\mathbf{X}_\ell) \right),
\\
\mathbf{X}_{\ell+1} = \mathbf{Z}_\ell + {\rm FFN}_\ell \left( {\rm LN}(\mathbf{Z}_\ell) \right),
\end{align}
where $\ell=0,\dots,L_{\rm tfm}-1$, ${\rm LN}(\cdot)$ denotes layer normalization and residual connections are applied in each block.


After $L_{\rm tfm}$ layers we obtain interference-aware embeddings
$\tilde{\mathbf X}
=
f_{\rm tfm}(\mathbf X_0)
\in\mathbb R^{K\times d_{\rm m}}$,
which are directly used for beam prediction.

By jointly processing all user tokens, the encoder learns beam selections that account for E2E under the MMSE-based loss in \eqref{Loss}, rather than optimizing each user independently.

\subsubsection{Scalable beam selection head with Gumbel--Softmax}

To determine analog beams for the $N_{\rm sub}=K$ subarrays, we employ a parameter-shared multihead prediction module. For each subarray $n$, a shared linear head maps the subarray corresponding feature vector $\tilde{\mathbf{x}}_{n} = \tilde{\bf X}(n, :)$ to logits over the far-field codebook $\mathcal F$ as $\mathbf{p}_n = f_{\rm head}(\tilde{\mathbf x}_n) \in \mathbb R^{N_q}, n=1,\dots,N_{\rm sub}$.
Parameter sharing across subarrays keeps the model size independent of $N_{\rm sub}$.

During training, we adopt a Gumbel--Softmax relaxation to obtain a differentiable approximation of the one-hot beam index
\begin{equation}
\boldsymbol{\pi}_n = {\rm softmax} \left( \frac{\mathbf{p}_n+\mathbf{v}_n}{\tau} \right) \in\mathbb R^{N_q},
\end{equation}
where $\mathbf{v}_n$ is i.i.d.\ Gumbel noise and $\tau$ is the temperature.

The analog beam of subarray $n$ is constructed as
\begin{equation}
\hat{\mathbf f}_n = \sum_{i=1}^{N_q} \pi_{n,i} \mathbf{a}_{N_a}(\theta_i) \in \mathbb C^{N_a \times 1},
\end{equation}
and the block-diagonal analog beamformer becomes $\hat{\mathbf F}_{\rm RF} = {\rm diag} \{\hat{\mathbf{f}}_1,\dots,\hat{\mathbf{f}}_{N_{\rm sub}}\}$. This relaxation enables backpropagation through the discrete beam selection under the loss \eqref{Loss}.
During inference, the beam index is obtained by $\hat{i}_n=\arg\max_i [\mathbf{p}_n]_i, \hat{\mathbf{f}}_n
=\mathbf{a}_{N_a}(\theta_{\hat{i}_n})$.

\section{Simulation Results}
The base station serves $K=8$ users, each consisting of $N_a=32$ antennas. The carrier frequency is set to 100 GHz. For each channel realization, $\alpha_{\ell, k}\sim\mathcal{CN}(0,1)$, $\sin\theta_{\ell, k}\sim\mathcal{U}[-1,1]$, and $r_{\ell, k}\sim\mathcal{U}[5, 200]$ m. Unless otherwise specified, in the proposed DL-IABT, $M=8$, and $\mathcal{F}$ contains $N_q=32$ beams. We train the network using AdamW with learning rate $5 \times 10^{-4}$, weight decay $2 \times 10^{-3}$, batch size 1024, dropout 0.3, and a cosine annealing scheduler. For DL-IABT, the dimensions of complex-valued sensing front-end are $[512, 256, 128]$, and 2-layer Transformer encoder is configured with $H=2$ attention heads.   

Several representative baselines are considered. \textbf{Ideal PCSI AO} assumes perfect CSI and applies alternating optimization to jointly design the analog and digital beamformers, serving as an upper performance bound. \textbf{Noisy NCSI AO} performs the same optimization using noisy channel observations. \textbf{Radix-4 BT} follows the hierarchical search strategy in \cite{zhang2019subarray}. An \textbf{MLP-based predictor} is also implemented to directly map sensing measurements to beam indices and is trained to maximize the sum rate. Finally, \textbf{Random beam selection} randomly chooses analog beams from the codebook. 

\begin{figure}[!t]
\centering
\includegraphics[width=2.5in]{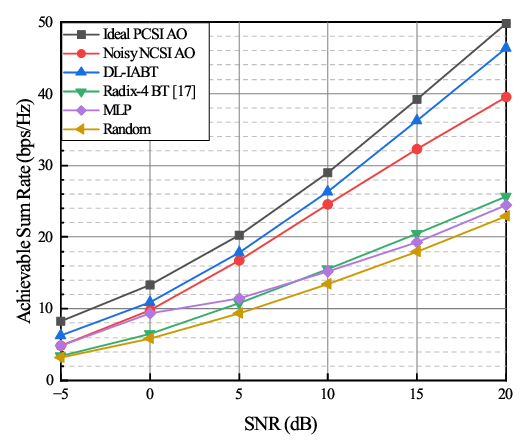}
\caption{Achievable sum rate versus SNR.}
\label{fig:SumRate}
\end{figure}

As shown in Fig. \ref{fig:SumRate}, the proposed DL-IABT scheme consistently approaches the ideal PCSI AO upper bound and significantly outperforms the noisy NCSI AO baseline. For instance, at 20 dB, the proposed method achieves 46.33 bps/Hz, which is close to the ideal result of 49.83 bps/Hz. In contrast, Radix-4 suffers noticeable performance loss since it does not explicitly account for E2E during beam selection. Although the MLP baseline optimizes the sum-rate objective, it processes the sensing features using MLP without explicitly modeling the interactions among users. Consequently, the network must implicitly learn the E2E structure from data, which is difficult in high-dimensional beam selection problems. Random beam selection provides the lowest performance as expected.

\begin{figure}[!t]
\centering
\includegraphics[width=2.5in]{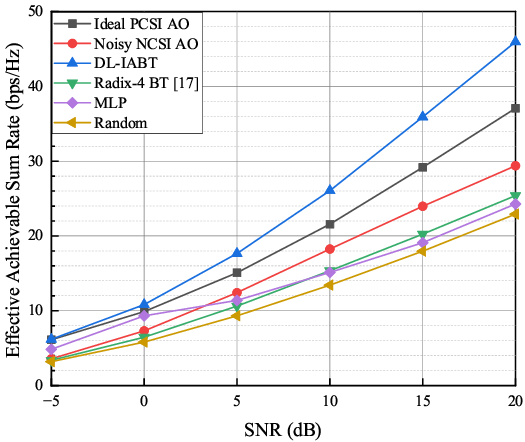}
\caption{Effective achievable sum rate versus SNR.}
\label{fig:EFSR}
\end{figure}

Fig. \ref{fig:EFSR} illustrates the effective achievable sum rate $R_e = (1-Mt/T_c) R_{\rm sum}$ versus SNR when the channel coherence time $T_c$ is 200 ms and each pilot transmission occupies $t=0.2$ ms. Compared with the achievable sum-rate results, the performance of linear beam searching schemes is noticeably reduced due to the pilot overhead. In particular, the ideal AO and noisy AO baselines suffer a significant degradation since $N_q$ pilot measurements are needed. In contrast, the proposed DL-IABT scheme requires only a small number of pilot observations and therefore achieves the highest effective rate across all SNR regimes. For instance, at 20 dB, the proposed method achieves 45.96 bps/Hz, which is substantially higher than 37.08 bps/Hz of ideal AO and 29.39 bps/Hz of noisy NCSI AO. These results demonstrate that the proposed framework effectively improves spectral efficiency by significantly reducing pilot overhead.

\begin{figure}[!t]
\centering
\includegraphics[width=2.5in]{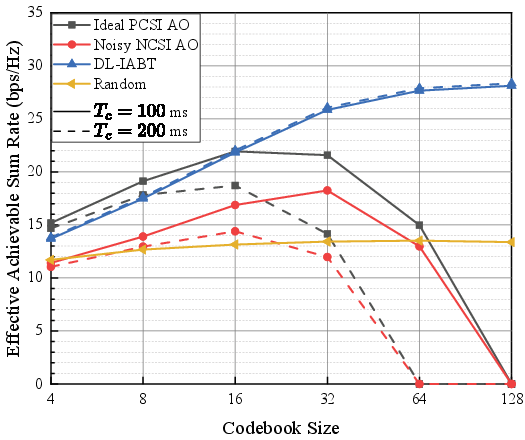}
\caption{Effective achievable sum rate versus codebook size.}
\label{fig:CodebookSize}
\end{figure}

Fig. \ref{fig:CodebookSize} shows the effective achievable sum rate versus the codebook size at ${\rm SNR=10\ dB}$. As the codebook size increases, the performance of the AO-based schemes first improves and then rapidly degrades due to the increasing pilot overhead required for beam search. In particular, when the codebook size becomes large, the training overhead exceeds the channel coherence time, resulting in 0 bps/Hz for both ideal and noisy AO. In contrast, the proposed DL-IABT scheme maintains stable performance as the codebook size increases since it predicts beam indices directly from a small number of pilot observations. Consequently, the proposed method achieves significantly higher effective spectral efficiency in large codebook scenarios, demonstrating its strong scalability with respect to beam resolution.

\section{Conclusion}
This letter studied pilot-limited multiuser BT for sub-connected XL-MIMO systems operating in mixed near- and far-field regimes. By exploiting a subarray-level approximation, we adopted a far-field codebook to represent subarray responses and reduce the near-field BT burden. To enable E2E learning under discrete codebook constraints, we derived a variant-MSE surrogate objective by eliminating the digital precoder through a closed-form MMSE solution based on KKT conditions, which implicitly captures E2E. Building on this objective, we proposed DL-IABT, which integrates a complex-valued sensing front-end, a shared complex-valued feature encoder, a Transformer-based multiuser predictor, and a scalable Gumbel--Softmax beam selection head. Simulations demonstrate that DL-IABT achieves a near-optimal sum-rate and, when pilot overhead is taken into account, delivers the highest effective throughput.



 
\bibliographystyle{IEEEtran}
\bibliography{IEEEabrv,References}

\vfill

\end{document}